\documentclass{article} 
\usepackage{MLDD_workshop_2023, times}

\usepackage{amsmath,amsfonts,bm}

\def\eqref#1{equation~\ref{#1}}

\def\1{\bm{1}}

\DeclareMathAlphabet{\mathsfit}{\encodingdefault}{\sfdefault}{m}{sl}
\SetMathAlphabet{\mathsfit}{bold}{\encodingdefault}{\sfdefault}{bx}{n}

\usepackage{hyperref}
\usepackage{url}
\usepackage{graphicx}
\usepackage{enumitem}
\usepackage{mathtools}
\usepackage{placeins} 

\title{Accurate Free Energy Estimations of Molecular Systems Via Flow-based Targeted Free Energy Perturbation}

\author{Soo Jung Lee\thanks{Equal contribution.}, Amr H. Mahmoud \footnotemark[1] \& Markus A. Lill \\
Department of Pharmaceutical Sciences\\
University of Basel\\
Klingelbergstrasse 50\\ 
4056 Basel, Switzerland\\
\texttt{markus.lill@unibas.ch} \\
}

\iclrfinalcopy 
\begin{document}

\maketitle
\begin{abstract}
The Targeted Free Energy Perturbation (TFEP) method aims to overcome the time-consuming and computer-intensive stratification process of standard methods for estimating the free energy difference between two states. To achieve this, TFEP uses a mapping function between the high-dimensional probability densities of these states. The bijectivity and invertibility of normalizing flow neural networks fulfill the requirements for serving as such a mapping function. Despite its theoretical potential for free energy calculations, TFEP has not yet been adopted in practice due to challenges in entropy correction, limitations in energy-based training, and mode collapse when learning density functions of larger systems with a high number of degrees of freedom. In this study, we expand flow-based TFEP to systems with variable number of atoms in the two states of consideration by exploring the theoretical basis of entropic contributions of dummy atoms, and validate our reasoning with analytical derivations for a model system containing coupled particles. We also extend the TFEP framework to handle systems of hybrid topology, propose auxiliary additions to improve the TFEP architecture, and demonstrate accurate predictions of relative free energy differences for large molecular systems. Our results provide the first practical application of the fast and accurate deep learning-based TFEP method for biomolecules and introduce it as a viable free energy estimation method within the context of drug design. 
\end{abstract}

\section{Introduction} \label{Introduction}

Free energy calculation methods such as Free Energy Perturbation (FEP) \citep{zwanzig1954high} are valuable in the field of Computer-aided Drug Design (CADD) to evaluate binding affinities between candidate compounds and receptors, or other biomolecular interactions \citep{Brown2010, pohorille2010good, de2016role}. Although useful in cases when experimental testing is unfeasible, high computational cost remains the current drawback for otherwise reliable and accurate in silico methods. For example, FEP calculations for the free energy difference between two thermodynamic states requires a multitude of stratified molecular dynamics (MD) simulations so that the distributions of the explored configuration space by these intermediates sufficiently overlap to achieve convergence (Section \ref{apx_fep}). 

Targeted Free Energy Perturbation (TFEP) has been presented as a potential alternative method for free energy calculation by applying a generalized FEP identity (Equation \ref{eq:fepidentity}) whereby an invertible, high-dimensional mapping function is used to transform a distribution of one system or state to another \citep{jarzynski2002targeted, hahn2009using}. The invertible and bijective normalizing flow neural network has been suggested as a solution to overcoming the difficulty of formulating the required complex mapping functions in order to achieve overlap of configuration space distributions, and its application for TFEP has previously been demonstrated on a growing soft sphere solute in a solvated box \citep{wirnsberger2020targeted}. 

Despite early success, several challenges remained that hindered immediate application of the TFEP method to larger systems with many more degrees of freedom (DOFs) which follow more complex density functions, such as biomolecules. In our study, we expand on the flow-based TFEP method by addressing the following points: 

\begin{itemize}
    \item The bijectivity of flow-based generative neural networks must be upheld by keeping the DOF consistent throughout, a critical point considering many biomolecular free energy studies are performed to compare systems of different sizes. We convert systems to hybrid topologies where deleted atoms are replaced by dummy atoms and inserted atoms are generated by geometric proposal engines. 

    \item The use of dummy atoms in flow-based TFEP for single free energy difference calculations ($\Delta F$), as opposed to dual free energy differences ($\Delta \Delta F$), leads to inaccuracies due to their entropic contributions. We first delineate this effect through analytical derivations using the coupled-particle toy system. We then propose two different methods of circumventing free energy estimation variance due to the presence of dummy atoms. The first method is incorporation of an auxiliary flow model to the TFEP flow architecture. The second method uses the thermodynamic cycle for relative free energy difference calculation to cancel out entropic effects resulting from dummy atoms.

    \item The difficulty in training a flow-based neural network rises concomitantly with increases in the DOF. A secondary benefit of adding the auxiliary flow is that it approximates a complex target probability density function which can then be set as prior for the free energy difference estimating bijector. The similarity between the prior and target densities reduces the scale and distance of transformation necessary by flow and impedes mode-seeking behavior that often leads to incorrect learning.

    \item We use our extended TFEP method to demonstrate successful prediction of free energy differences for two different tasks commonly studied in chemical biology and drug discovery: Computation of hydration free energy and free energy differences of protein stability due to mutagenesis. The results are comparable in accuracy to established computational methods at significantly reduced computational costs due to circumvention of long simulations. 

\end{itemize}

\begin{figure}[ht]
\begin{center}
\includegraphics[width=\textwidth]{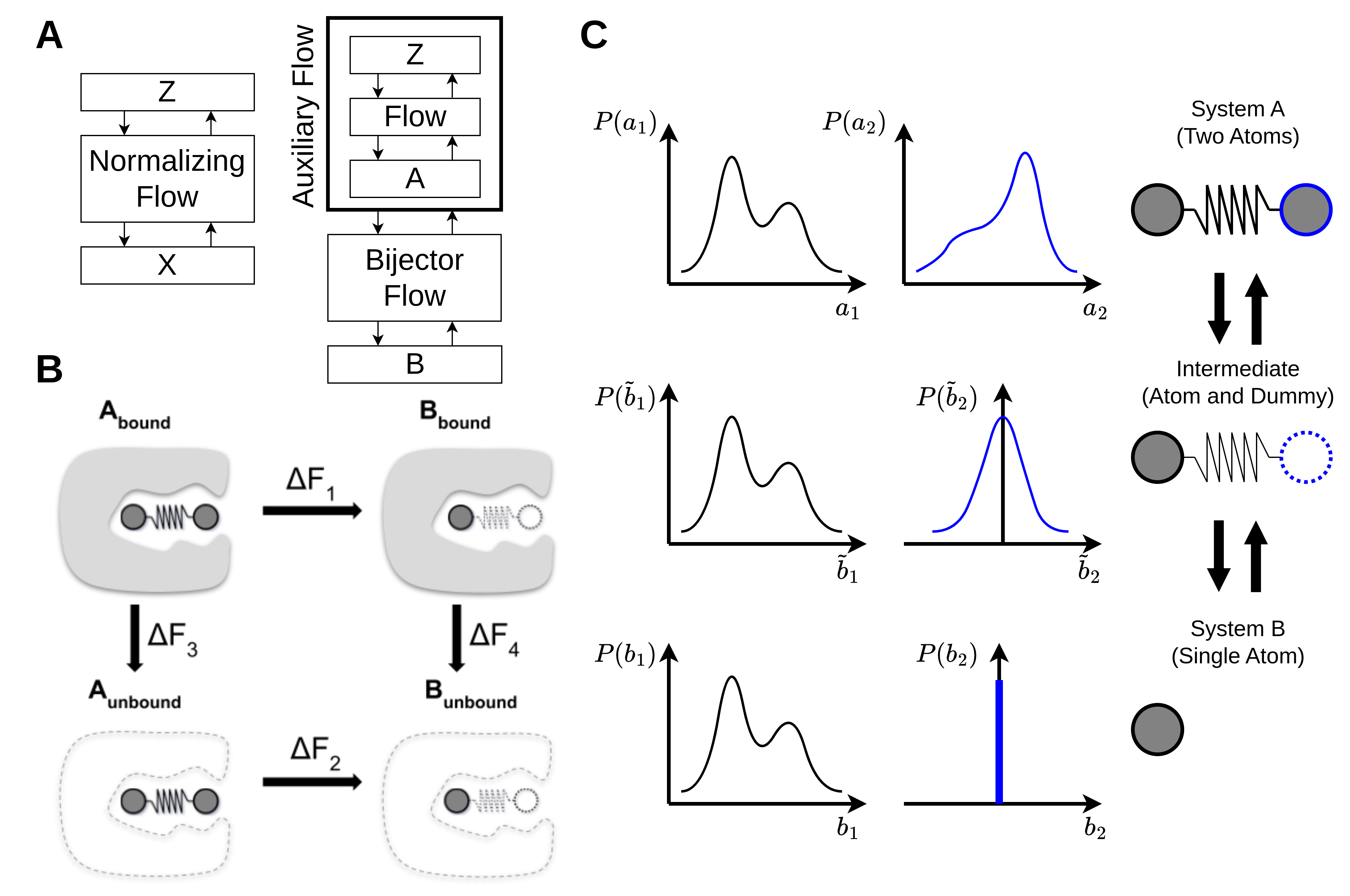}
\end{center}
\caption{Schematics for the toy system example. (A) Comparison of a standard normalizing flow (left) and an auxiliary flow stacked on a bijector flow (right). Normalizing flows conventionally learn mapping between a data distribution $\hat p_{X}$ for data space $X$ and a prior distribution $p_{Z}$ (typically Gaussian) for reference space $Z$. (B) Schematic of the thermodynamic cycle for the coupled particle system. Dotted line particles represent dummy atoms. (C) Scheme representing the process of atom annihilation and its corresponding probability density.  The probability density of the annihilated particle is transformed to a Gaussian distribution (intermediate state, with dummy atom). The theoretical mapping of this density to a Dirac delta function cannot be modeled by flow-based methods but is computed by analytical means.}
\label{fig:toy}
\end{figure}

\section{Related Works}
\subsection{Targeted Free Energy Perturbations}
TFEP is an elegant approach to free energy difference estimation devised by \citet{jarzynski2002targeted}, extending Zwanzig's FEP identity function \citep{zwanzig1954high}. For two thermodynamic states $A$ and $B$, the relationship between the true free energy difference $\Delta F$ of the two states can be recovered using a generalized estimator $\Phi$ instead of potential energy differences $\Delta U$:
\begin{equation} \label{eq:fepidentity}
    e^{-\beta \Delta F} = \mathbb{E}_A\left[ e^{-\beta \Phi_{A\rightarrow A'}} \right],
\end{equation}
where $\beta$ denotes the thermodynamic beta and $A'$ the new proposal distribution so that for configurations $\mathbf{x} \sim A$, the mapping is $M(\mathbf{x}) \sim A'$. In the forward transfer mapping $A$ to $B$, expectations are taken with respect to the equilibrium density of $A$, $\rho_A \propto e^{-\beta U_A}$ (similar for $B$ when mapping the reverse from $B$ to $A$).

The neural network perfoming the role of TFEP mapping function $M : A \rightarrow A'$ must be invertible so that $M^{-1} : B \rightarrow B'$. The generalized energy differences for forward and reverse directions are defined as the difference in potential energies between target and base, from where the log determinant of the Jacobian ($J$) associated with the respective map direction has been subtracted:
\begin{align} \label{eq:phi}
    \Phi_F ( \mathbf{x} ) &= U_B ( M ( \mathbf{x}) ) - U_A ( \mathbf{x} ) - \beta^{- 1} \log {\left| {J_M( \mathbf{x} )} \right|} \nonumber\\
    \Phi_R ( \mathbf{x} ) &= U_A ( M^{- 1} (\mathbf{x} ) - U_B ( \mathbf{x}) - \beta^{- 1} \log{\left| {J_{M^{- 1}} ( \mathbf{x} )} \right|}.
\end{align}
Additional details are provided in Section \ref{apx_tfep}.

TFEP harnesses a path-independent mapping function to overlap the configuration space distribution between reference and target state. With a correctly formulated mapping function in place, convergence of the free energy differences is instantaneous and can significantly accelerate free energy predictions compared to conventional methods dependent on simulations which require a lengthy sequence of energy calculations per stratification step. The bottleneck thus far has been the difficulty in formulating such a complex mapping function, for which normalizing flow neural networks have recently been presented as a solution \citet{wirnsberger2020targeted, rizzi2021targeted, falkner2022conditioning}. Normalizing flows rely on identical dimensionality between reference and target spaces for bijector transforms in order to perform exact density estimations, a point that is discussed further in Section \ref{section_2d_entropy}. 

\subsection{Normalizing Flows and Boltzmann Generators} 
\label{section_nf_bg}
The architecture of the mapping function $M$ is crucial for accurate free energy prediction using the TFEP method because convergence must be reached based on a finite number of samples obtained from the two end states $A$ and $B$. Previous reports exhibit the effectiveness of normalizing flows \citep{Papamakarios2019, Rezende2020} in free energy calculations \citep{ding2021computing, ding2021deepbar, Noe_2019}, which are transformations of variables $z$ from a known prior density such as a Normal distribution to those from a more complex distribution, $x = f(z)$. Boltzmann Generators are a type of flow-based generative model that learns such diffeomorphisms for molecular structures with many DOFs \citep{Noe_2019}. Additional details are provided in Section \ref{apx_bg}. 

\section{Entropic Contributions of Dummy Atoms}
\label{section_2d_entropy}

As reported by \citet{wirnsberger2020targeted}, normalizing flows satisfy requirements for mapping function implementation in TFEP in that they are bijective, allow for efficient computation of the inverse and the Jacobian determinant, and are highly flexible. The significance of bijectivity is that although DOFs need to stay conserved for flow-based TFEP, the systems being compared do not always share the same number of atoms for many free energy questions posed in computational chemistry. To preserve the same dimensionality throughout the bijective transformation layers of normalizing flow, dummy atoms can be used as placeholders and the systems can be represented with a hybrid topology. 

We designed a toy system where states $A$ and $B$ differ by a single particle to test our treatment of dummy atoms within the TFEP method, and validate that the differences in the Jacobian term are equivalent to the analytically derived entropy of a Normal distribution function. If a transfer mapping function $M:A \rightarrow A'$ with target $B$ is optimally formulated such that $A' = B$, the Jacobian term from the TFEP generalized estimator $\log {\left| {J_M( \mathbf{x} )} \right|}$ represents the entropy difference between $A$ and $B$.

To numerically derive the entropic contribution by dummy atoms, we use a 2-dimensional coupled-harmonic particle system (Figure \ref{fig:toy}B, C), defined by the energy functions provided in Section \ref{apx_coupled}. These potential functions confine movement of the particles to a predefined range of coordinate space to emulate a physical harmonic bond term, and represent biological ligands in bound and unbound states. The objective is to analytically solve for the explicit entropy contribution to free energy difference predictions by the dummy particle that replaces the deleted atom. 

The model we use for TFEP free energy estimation is a combination of two stacked flows as shown in Figure \ref{fig:toy}A. We construct an auxiliary flow as an RQNSF-based BG (\citet{durkan2019}, Section \ref{apx_rqnsf}) of 4 affine coupling layers with alternating even and odd binary masking using the nflows library \citep{nflows}. Supposing we perform a transfer mapping of $B$ to $A$, this auxiliary flow is foremost trained as a density estimator for $A$ (120 epochs, 4096 batch size, $5e^{-4}$ learning rate) by maximum likelihood. The prior is set as a Normal distribution function and the target as $A$. 

The auxiliary flow is stacked on top of a second bijector flow, which is the mapping function between $A$ and $B$ and responsible for the free energy difference estimation between the two systems. The bijector is constructed and trained similarly to the auxiliary flow, the only difference being that the prior is set as the estimated density of the auxiliary flow (the learned $A$), and the target is $B$. Training is conducted via maximum likelihood so that samples from $B$ are transformed such that the generated distribution $B'$ has maximized overlap in the domain of the distribution of the auxiliary flow-estimated $A$. Finally, the TFEP loss function is used to evaluate the transformation. 

The advantage of using a dual flow is that the learned prior and target of the bijector occupy distributions that are more similar than a normalizing flow with a noise prior, heightening the efficiency and accuracy of training (Figure \ref{si_fig:2d_results}). Invertible transformations between similar distributions are more straightforward and easier to find because they preserve the general structure of the distributions and can map similar regions to each other. This method avoids increased complexity in the transformation that can lead to difficulty in optimization, risk of overfitting, intractability in computation, and burdens in memory requirements. 

In contrast, training a bijector alone with no auxiliary density estimate would be driven by minimization of Kullback-Leibler (KL) divergence, which leads to mode-seeking behavior. For larger systems that follow highly complex and multimodal density functions, this behavior leads to difficulties in convergence and ultimately inaccurate free energy difference predictions. 

\begin{table}[htbp]
\caption{Entropy Effects on Single Free Energy Differences}
\label{table_2d}
\begin{center}
    \begin{tabular}{|c|cc|cc|}
\hline
    & \multicolumn{2}{c|}{Ground truth}           & \multicolumn{2}{c|}{Flow}            \\ \hline
    & \multicolumn{1}{c|}{Bound}  & Unbound & \multicolumn{1}{c|}{Bound} & Unbound \\ \hline
\multicolumn{1}{|r|}{$\Delta F$}  & \multicolumn{1}{c|}{-1.853} & -0.470  & \multicolumn{1}{c|}{-3.300} & -1.950   \\ \hline
\multicolumn{1}{|r|}{$\Delta \Delta F$} & \multicolumn{2}{c|}{-1.383}           & \multicolumn{2}{c|}{-1.350}           \\ \hline
\end{tabular}
\end{center}
\end{table}

As shown in Table \ref{table_2d}, we obtained single free energy differences ($\Delta F$) for atom deletion between prior ($A$, coupled two-particle system) and target ($B$, single-particle system) using the dual flow method with TFEP. In the bound state, the difference in $\Delta F$ between ground truth and flow is 1.447, and that for unbound state is 1.480. These values are indeed comparable with the known entropy of a Normal distribution, $S = \ln{\sqrt{2\pi}}  + \frac{1}{2} = 1.4189$. We are therefore able to demonstrate with our coupled particle deletion example that a dummy atom defined by a Gaussian will contribute entropically to single free energy difference estimations, and it is crucial to take this into consideration when using the TFEP method for free energy studies of molecular systems with changes in DOFs. 

\section{Construction of Hybrid Topology}
\label{section_hybrid}
\begin{figure}[ht]
\begin{center}
\includegraphics[width=\textwidth]{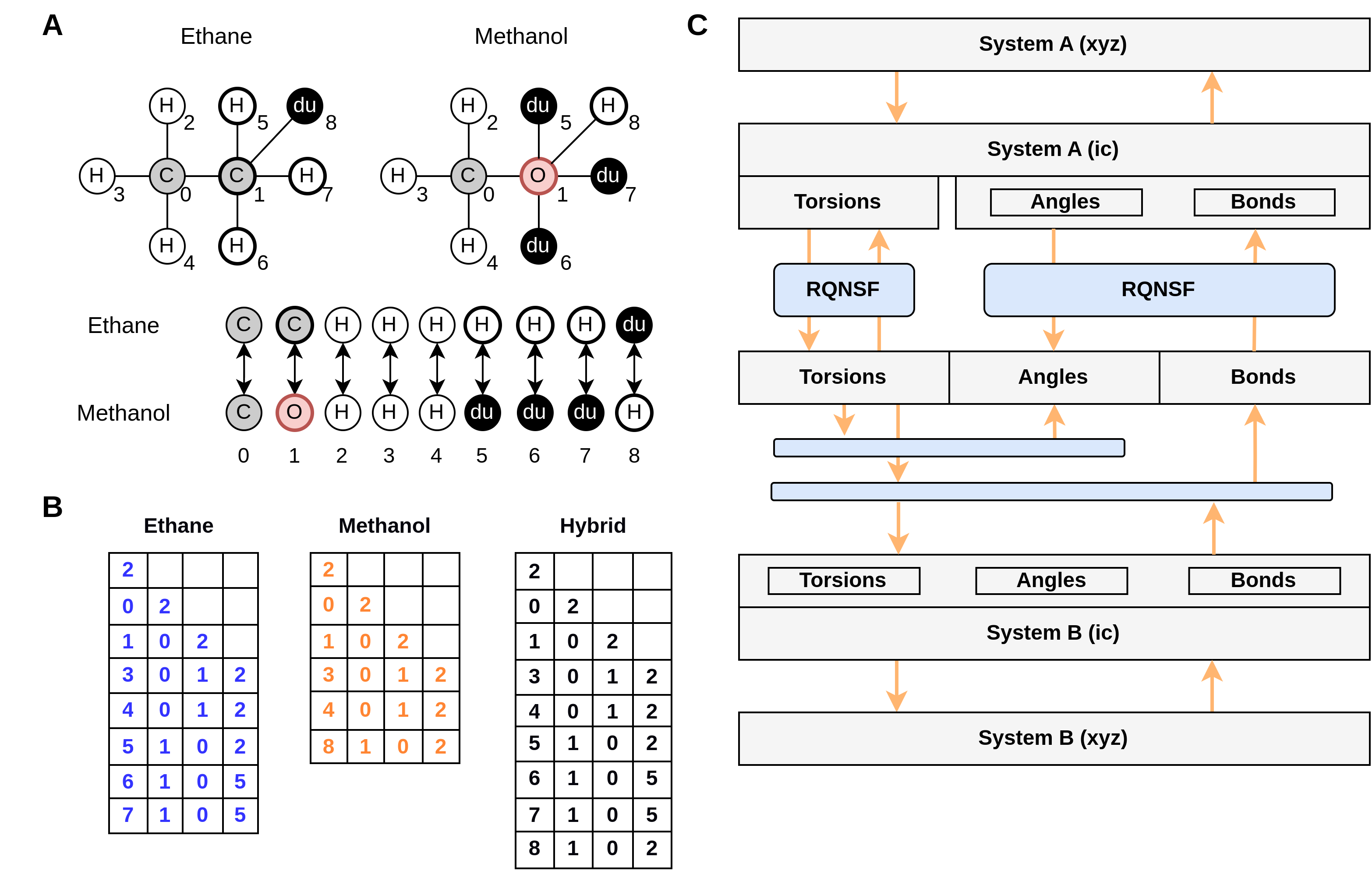}
\end{center}
\caption{Flow-based TFEP model implementations. (A) Hybrid topology for ethane and methanol with dummy atoms (du) inserted for consistent DOFs in bijective transformation via flow models. Environment atoms are numbered 0, 2-4, core atoms are numbered 1, and unique atoms are numbered 5-8 and marked with a bold outline. Black bidirectional arrows indicate the mapping of each atom to its corresponding atom in the other system.  (B) $Z$-matrices for ethane, methanol, and the hybrid systems. Atom numbering identical to (A). (C) Schematic summarizing the flow architecture that transforms between base System A and target System B. RQNSF coupling layers are represented as blue blocks, Cartesian coordinates are labeled `xyz', and internal coordinates are `ic'.}
\label{fig:architecture}
\end{figure}

Relative free energy difference studies in drug discovery often involve end states that have different numbers of atoms, or DOFs. To preserve the bijective nature of normalizing flow for the TFEP method and simultaneously allow the method to be applicable for systems of different size and topology, we implement a hybrid topology approach. Hybrid topologies fuse two systems and map their corresponding atoms to each other (Figure \ref{fig:architecture}A). In this section, we describe the particle types in hybrid topology, the setup for traditional relative free energy calculations by FEP approach (RFEP), and how it is integrated in our proposed deep learning model. 

Particles in a hybrid topology are assigned to one of four groups: environment, core, unique old, and unique new. Suppose we are estimating free energy differences between Molecule A and Molecule B. All atoms of common residues in A and B are environment atoms, while common atoms on the differing residue are designated core atoms (e.g. the C-alpha atom of the mutated residue in the Trpcage W6F study, Section \ref{section_trpcage}). When considering a transformation in the direction from A to B, unique atoms for A would be considered unique old and those for B would be considered unique new. 

 RFEP can be divided into three stages: First, a hybrid 2D topology is generated from single topologies of Molecules A and B. Second, the topology and coordinates of Molecule A are used as input to initiate a hybrid 3D system. Third, atoms are generated using a geometric proposal engine (Additional details, Section \ref{apx_rfep}).

For our study, we follow strategies implemented in the Perses framework \citep{rufa_dominic_a_2022_6328265} that use a variant of maximum common substructure algorithm from OEChem TK (OpenEye Scientific Software, Santa Fe, NM) and the force constant and equilibrium state of dummy atoms.

\section{Architecture and Method Design}
Variations of normalizing flow networks have been published, where some are permutation equivariant but lacking rotation-translation equivariance \citep{wirnsberger2020targeted} and others exhibit E(n) equivariance at the expense of learning instability \citep{Satorras2021}. We base our implementation (Fig. \ref{fig:architecture}C) on the BG concept \citep{Noe_2019} with the following minor adjustments: 
\begin{enumerate}[label=(\roman*)]
    \item We opt for a rotation-translation invariant solution by leveraging the space of internal coordinates as a product of hypertori and compact intervals, similar to previous literature \citep{Noe_2019, koehler2021, invernizzi2022skipping}, and bookend the bijector flow with layers that transform Cartesian to internal coordinates or vice versa. These are non-trainable layers that transform Euclidean coordinates $\boldsymbol{x} \in \mathbb{R}^{n \times 3}$ into bonds $\boldsymbol{d} \in\left[a_1, b_1\right] \times$ $\ldots \times\left[a_{n-1}, b_{n-1}\right]$, angles $\boldsymbol{\alpha} \in[0, \pi]^{n-2}$, and torsions $\boldsymbol{\tau} \in \mathbb{T}^{n-3}$. The internal coordinate transformation is performed by rational quadratic spline flows \citep{durkan2019}. The model learns a joint distribution $p(\boldsymbol{d}, \boldsymbol{\alpha}, \boldsymbol{\tau})$ on the topological space $\mathcal{X}_{\mathrm{IC}}:=\mathbb{I}^{2 n-3} \times \mathbb{T}^{n-3}$, where the closed unit interval is $\mathbb{I}=[0,1]$.
    \item We set priors as nontrivial probability distributions that are sampled via MD. MD simulations are performed using the hybrid topology for each end state (lambda 0 and 1) in Cartesian coordinates from which we obtain a representative set of configurations for training in the forward ($A$ to $B$) or reverse direction ($B$ to $A$). Due to the similarity between the prior and the target, the learning by energy scheme (\citet{Noe_2019}, Section \ref{apx_bg}) is much smoother and we are able to train at fast speed by using a simple flow architecture with only a few coupling layers.
\end{enumerate}

Normalizing flows are capable of tractable transformations, but they are not necessarily expressive enough to capture any and all complex mapping from a naive prior to a high-dimensional, complex target distribution. Training a BG in practice may also be difficult because energy-based training is characterized by mode-seeking behavior and struggles to converge reliably when the target distribution is multimodal. Our method overcomes this by designing a mapping between similar density functions, which makes the transform more straightforward and enables the model to train quickly and accurately for molecular systems. 

\subsection{Cancellation of Dummy Entropy}
\label{section_cancel}
For end states $A$ and $B$ that represent the initial Molecule A ($\lambda$=0, where $\lambda$ represents an alchemical coupling parameter) and final Molecule B ($\lambda$=1), we perform simulations using a hybrid topology as described in Section \ref{section_hybrid}, where dummy atoms are either turned on or annihilated during the transformation depending on which end state molecule they represent. Dummy atoms are connected to the common core through alchemical bonded force field terms, which can skew energies calculated for the end states if they are placed with disregard to physical context. Additionally, the DOFs of dummy atoms are included in the mapping between $A$ and $B$ and contribute entropically to the free energy difference.  

The first method of balancing these energetic and entropic contributions by implementing an auxiliary flow as a density estimator for the free energy predicting bijector has been previously demonstrated using a toy system in Section \ref{section_2d_entropy}. As a second approach, we take advantage of the dummy atoms decoupling from the core environment with regards to their nonbonded interactions in their annihiliated state. This causes the configurational samples of the dummy atoms and their energetic evaluation to be independent of the core and environment. Moreover, these independent energetic and entropic contributions of the dummy atoms cancel each other out in parallel legs of thermodynamic cycles. Dummy atoms in the solvated and vacuum states cancel each other out for solvation free energy calculations, those in folded and unfolded states do so for mutation free energy calculations, and those in bound and unbound complexes also for relative free energies of binding. As a result, it is unnecessary to correct for the energetic and entropic contributions of annihilated dummy atoms for single free energy differences ($\Delta F$) of each transformation leg separately. Instead, we directly compute the dual free energy difference ($\Delta \Delta F$) from the full thermodynamic cycle. 

\begin{figure}[htbp]
\begin{center}
\includegraphics[width=\textwidth]{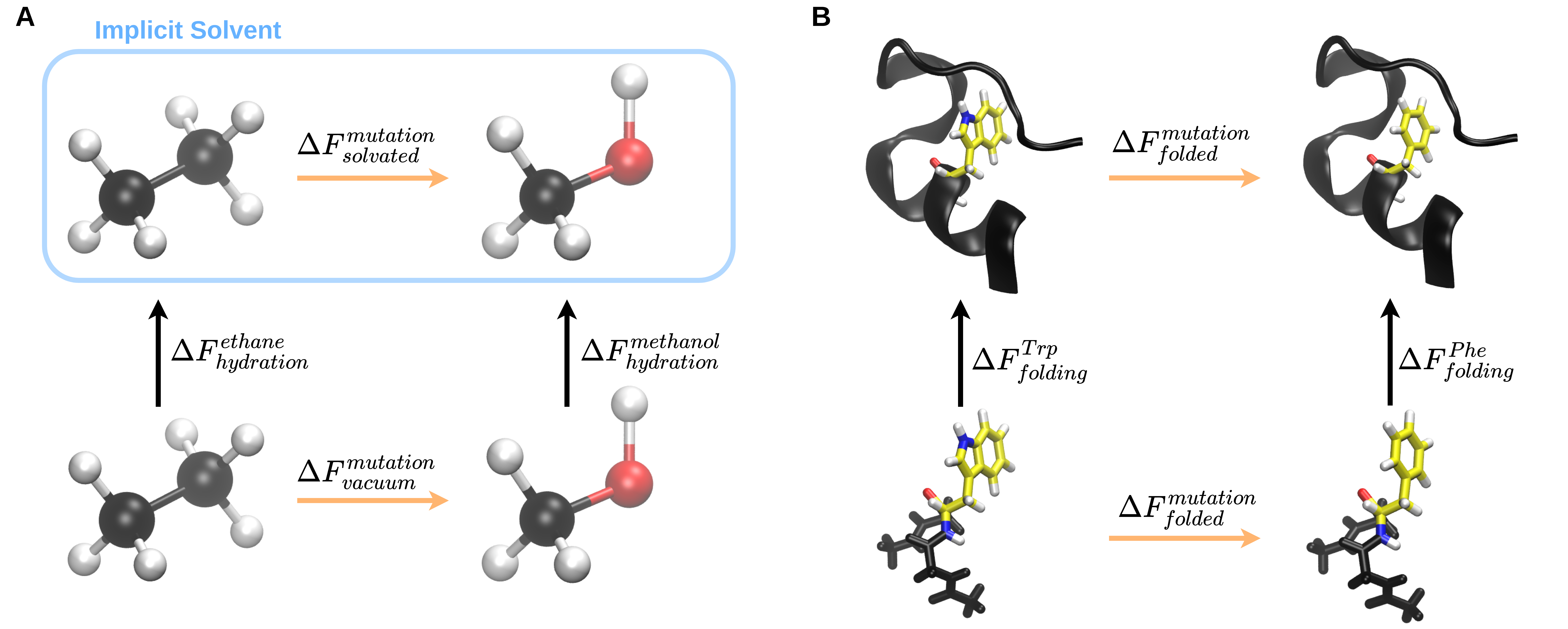}
\end{center}
\caption{Thermodynamic cycles for application example systems. (A) Thermodynamic cycle for mutation between ethane and methanol. (B) Thermodynamic cycle for W6F residue mutation in Trp-cage miniprotein. Legs of the thermodynamic cycle where the flow-based TFEP mapping function is used are marked orange.}
\label{fig:thermocycle}
\end{figure}

\section{Application on Small Molecule Mutation}
The energy difference when a molecule transitions from vacuum to a solvated state at constant temperature and pressure is represented by solvation free energy, $\Delta F^{\text {solvation}}$ (or hydration free energy in cases where the solvent is water). Solvation free energy difference estimations are pertinent to computational chemistry and drug design and traditionally require stratified, alchemical methods (Section \ref{apx_fe_calculation}). Commonly, a series of nonphysical intermediates parameterized by the alchemical coupling parameter $\lambda$ are simulated to ultimately transfer a solute from gas phase to solvent or vice versa.

We test our TFEP method on prediction of hydration free energy difference between ethane and methanol. From the thermodynamic cycle in Figure \ref{fig:thermocycle}A, we see that $\Delta \Delta F_{hydration} = \Delta F^{methanol}_{hydration} - \Delta F^{ethane}_{hydration} = \Delta F^{mutation}_{solvated} - \Delta F^{mutation}_{vacuum}$. In this section, we investigate two pathways. The first is the solvation pathway adapted from \citet{duarte2017approaches}, where ethane and methanol are respectively flow transformed from water to vacuum or vice versa. The energy functions that calculate potential energy for each end state include terms and parameters for either vacuum or implicit solvent. 

\begin{table}[ht]
\label{table_ethane_solv}
\caption{Free energy difference estimation from solvation legs.}
\begin{center}
\begin{tabular}{lllll}
\hline
         & Experimental & Vac $\rightarrow$ Water & Water $\rightarrow$ Vac & Gromacs TI \\ \hline
Ethane    & 1.83         & 1.89                     & -1.91                    & 2.46       \\ 
Methanol  & -5.10        & -4.75                    & 4.71                     & -3.49  
\\\hline
Hydration $\Delta \Delta F$ (kcal/mol) & -6.93        & -6.64                   &         & -5.95  
\\
Dehydration $\Delta \Delta F$ (kcal/mol) &         &                  &  6.62  &  
\\ \hline
\end{tabular}
\end{center}
\end{table}

For the second mutation pathway experiment, we use the hybrid $z$-matrix topology scheme constructed by merging the $z$-matrices of ethane and methanol as shown in Figure \ref{fig:architecture}B. We compare the flow-based free energy estimations (NF) with reference methods thermodynamic integration (SOMD-TI) and Multiple Bennett Acceptance Ratio (SOMD-MBAR) (Section \ref{apx_fe_calculation}) following alchemical free energy calculation molecular dynamics simulations run by the OpenMM-interfacing SOMD engine \cite{loeffler2015fesetup, hedges2019biosimspace}. NF results are comparable to other methods as shown in Table \ref{table:ethane_ref}. The loss profile of the transformation of ethane to methanol in solvated and vacuum states and the solvation free energy is given in Figure \ref{si_fig:solvation_learning}. 

\begin{table}[ht]
\caption{Comparison with alternative free energy calculation methods.}
\begin{center}
\begin{tabular}{llll}
\hline
         & NF & SOMD-TI  & SOMD-MBAR\\ \hline
Ethane $\rightarrow$ Methanol (kcal/mol)  & -5.75         & -6.02                     & -6.16 \\ \hline
\end{tabular}
\end{center}
\label{table:ethane_ref}
\end{table}

\section{Application on Protein Residue Mutation}
\label{section_trpcage}
Trp-cage or TC5b \citep{neidigh2002designing} is a 20-residue (NLYIQWLKDGGPSSGRPPPS) miniprotein with a two-state folding mechanism that has been featured extensively in energy and conformation change studies \citep{simmerling2002all, snow2002absolute, chowdhury2003ab, nikiforovich2003possible, pitera2003understanding, zhou2003trp, ding2005simple, linhananta2005equilibrium, sidky2019high, juraszek2006sampling}. The key stabilizing components are the six residues making up the buried hydrophobic core (TYR3, TRP6, LEU7, GLY11, PRO12, PRO19), four prolines that minimize  $\Delta S_U$, and a salt bridge interaction.

Many single-site mutation studies have been performed on Trp-cage, both empirically and computationally. Among these we use W6F, which reportedly leads to completely unfolding in water with destabilization of $\Delta \Delta G_F = 12.5 \pm 0.6$ kJ/mol \citep{barua2008trp}.  

Tripeptides and mutation structure files were created and prepared using \citet{noauthor_schrodinger_nodate} Maestro  (wildtype PDB ID: 1L2Y). Explicitly solvated structures were prepared using the pdb4amber tool from the AmberTools MD package \citep{case2020amber} and MD simulations were run using OpenMM \citep{eastman2017openmm}. Frames from a 40 ns simulation were collected at equal intervals and used as starting states for multiple short (1 ns) simulations. All spawned simulations were concatenated as a single 40 ns trajectory of 400,000 frames to be used as training input structures to ensure statistically converged results \citep{genheden2010obtain}. For all simulations, nonbonded cutoff was 0.9 nm, temperature 300 K, timestep 1fs, and a Langevin dynamics integrator was used with 1 ps$^{-1}$ friction coefficient. Solvent molecules were removed and potential energy was calculated implicitly prior to network training (\citet{paschek2007replica}, Section \ref{section_discussion}).

We obtained flow-based TFEP free energy difference estimations for W6F mutation (NF) and compared the results with predictions from a non-equilibrium switching method implemented by open source software pmx \citep{seeliger2010protein, gapsys2015pmx}. We also compared estimations from equilibrium replica-exchange via lambda hopping (FEP-REMD) as implemented by Schr{\"o}dinger \citep{wang2015accurate}. 

\begin{table}[ht]
\caption{Results for Trp-cage W6F mutation free energy estimation.}
\begin{center}
\begin{tabular}{lllll}
\hline
         & Experimental & NF & pmx  & FEP-REMD (MBAR) \\ \hline
W6F (kcal/mol)  & 2.98  & 3.18 & 3.48  & 3.19 \\ \hline
\end{tabular}
\end{center}
\label{table:trpcage}
\end{table}

As shown in Table \ref{table:trpcage}, results for the flow-based TFEP method are comparable with both experimental and computational free energy difference values.

\section{Conclusion and Discussion} \label{section_discussion}
In this work, we introduce a range of contributions to the concept of flow-based free energy difference estimations and related physical problems. The objective of the proposed approach is to provide an alternative computational method of free energy estimation that performs at accelerated speeds compared to conventional methods by eliminating the need for a multitude of computationally expensive, stratified MD simulations. 

To summarize, we first acknowledge dummy atoms are necessary for preserving bijector dimensionality and these dummies should not influence the free energy calculations. We then analytically derive the Jacobian contribution in the TFEP generalized estimator by these dummy particles using a toy system.

In the next section, we propose integration of a hybrid topology scheme to the TFEP method as an efficient way of preserving dimensionality for molecular systems in normalizing flows. We also discriminate our method from normalizing flow-based BGs \citep{Noe_2019, invernizzi2022skipping, mahmoud2022accurate} by mapping between similar prior and target instead of between a noise prior and complex target, enhancing efficiency and the capacity to train larger systems. 

We also provide two approaches to ensure that deep-TFEP free energy difference estimations do not suffer from variance due to dummy atoms. The first is an architectural solution, stacking an auxiliary flow density estimator on the bijector as a prior. The second is by taking advantage of double free energy difference settings, under assumption that free energy differences for parallel legs of the thermodynamic cycle will contain the same entropic contributions that cancel out. A caveat to note is that this assumption may not hold true in all cases \cite{boresch1996jacobian, fleck2021dummy}, indicating a need for further study. 

Regarding training data for our deep generative models, we used Generalized Born (GB)-solvated data samples. Although better agreement can be achieved with experiments using explicit solvent, the choice for implicit solvation was deliberate due to challenges in modelling ensemble distributions that included explicit solvent molecules. Furthermore, inclusion of waters greatly increases the system size (DOFs) and also introduces permutation symmetry. Future studies on this topic may attempt a combination of promising developments in flow-based generative models for large systems with permutational symmetry \citep{Noe_2019, wirnsberger2020targeted, kohler2023rigid} with the current TFEP approach. Another possible direction for improving the current approach is to enhance physical accuracy by integrating deep learning potentials that implicitly account for explicit water \citep{chen2021machine,mahmoud2020elucidating,ghanbarpour2020instantaneous}. 

\subsubsection*{Acknowledgments}
This study was financially supported by funding from the Swiss National Science Foundation (Project number: 310030 197629).

\bibliography{iclr2023_conference}
\bibliographystyle{iclr2023_conference}

\appendix
\section{Appendix}
\setcounter{figure}{0}                   
\renewcommand\thefigure{S\arabic{figure}}
\subsection{Free Energy Perturbation}
\label{apx_fep}
 The pairwise free energy difference for the transfer between neighboring intermediate states are individually calculated and summed to obtain the overall free energy change:
\begin{align*}
\Delta F &= F_B - F_A \\
&= (F_B - F_N) + (F_N - F_{N-1}) + ... + (F_1 - F_A)\\
&= -k_B T \sum^N_{i=0} \ln  \left< \exp \left(   - \frac{U_{i+1} - U_i}{k_B T} \right) \right>_i
\end{align*}
The flow-based TFEP method allows us to bypass the computational expense of simulating numerous intermediate states $i$ and accelerate computation time without compromising the accuracy of free energy difference estimation.

\subsection{Deep Mapping in Targeted Free Energy Perturbation} 
\label{apx_tfep}
The relationship between the Kullback-Leibler Divergence loss function and the generalized estimator has been reformulated by \citet{wirnsberger2020targeted}:
\begin{equation*}
    D_{KL} \left[ \rho_{A'} || \rho_{B} \right] = \beta ( \mathbb{E}_A [\Phi_F] - \Delta F ).
\end{equation*}

When both $\Phi_F$ and $\Phi_R$ are available, the Bennett Acceptance Ratio (BAR) can be used to predict statistically accurate free energy:
\begin{equation*}
    \mathbb{E}_A[f(\beta(\Phi_F - \Delta F))] = \mathbb{E}_B[f(\beta(\Phi_R - \Delta F))],
\end{equation*}
with $f(x) = 1/(1+e^x)$.

While not directly related to the free energy calculations, the TFEP loss function has also been recently used to develop LREX to skip the replica-exchange ladder \citep{invernizzi2022skipping}.

\subsection{Boltzmann Generators} \label{apx_bg}
\begin{figure}[ht] 
\begin{centering}
\includegraphics[width=0.8\textwidth]{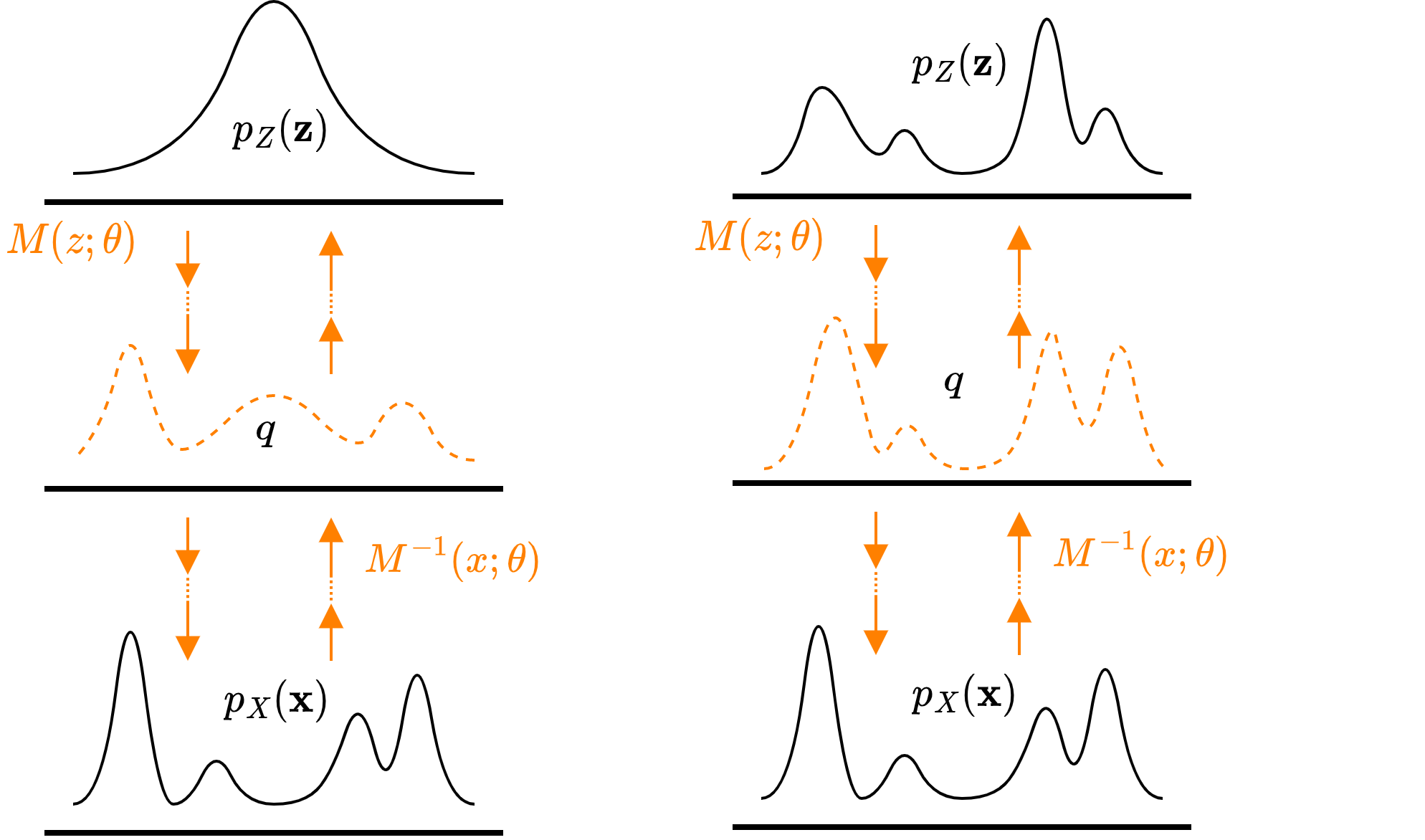}
\par\end{centering}
\caption{\textbf{Schematic for distribution learning in Boltzmann Generators.} The flow-based model learns the unique distribution for a protein system using a sequence of bijective, invertible functions, here depicted as dot-connected orange arrows. The exact prior and exact target are $p_{Z}(\mathbf{z})$ and $p_{X}(\mathbf{x}) \propto\exp(-u(\mathbf{x}))$. The BG-generated distribution is $q$, so that $q_{X}(\mathbf{x})=M(\mathbf{z})$ and $q_{Z}(\mathbf{z})=M^{-1}(\mathbf{x})$. The prior $p_{Z}(\mathbf{z})$ is typically a Gaussian for BGs (left). In our method, we set an auxiliary density estimator as prior instead of a Gaussian (right).}
\label{si_fig:bg}
\end{figure}

The multimodal complexity of energy function-dependent Boltzmann distributions requires a generative model that is capable of non-parametric density estimations. Flow-based Boltzmann Generators (BGs) have been reported to successfully obtain configuration samples of small molecular structures (up to 892 atoms) from the Boltzmann distribution $e^{-U(\mathbf{x})}$ following a training protocol based on valid reference structures from molecular dynamics (MD) and the energy function $U(\mathbf{x})$ \cite{Noe_2019}. BGs learn invertible coordinate transformations of variables $\mathbf{x} \sim p_{X}(\mathbf{x})$, configuration states which have high Boltzmann probability, from the latent space variables $\mathbf{z} \sim p_{Z}(\mathbf{z})$. The latent space distribution follows a simple Gaussian for easy backward propagation derivative calculations, same as other NFs (Figure \ref{si_fig:bg}, left). Each latent variable $\mathbf{z}$ from this fixed and prescribed probability distribution function belongs to a unique conditional distribution that is learnable by parameters $\theta$. The task of the BG network is therefore to learn $\theta$ for the transformations:
\begin{align} \label{eq:transform}
\mathbf{x}=M(\mathbf{z}; \theta), \:\:\:\:\:\: \mathbf{z}=M^{-1}(\mathbf{x}; \theta),
\end{align}

BGs are tailored for generating states of molecular systems because not only are the models trained using energetically and structurally valid states obtained through MD simulations (\emph{example-based training}), they are also trained on the energy function so that the target distribution $p_{X}(\mathbf{x}) \propto e^{-U(\mathbf{x})}$ (\emph{energy-based training}). Once $q$ has been learned and optimized, structures that do not violate the Boltzmann distribution can be generated from the trained model in one-shot.

\newpage
\subsection{Energy Function Definitions for the Coupled Particle System.}
\label{apx_coupled}
\textbf{Bound State}
\begin{alignat*}{4}
    &\textbf{System A} \:\:\:\: &&V_1(x_1, x_2) &&= 0.5\cdot k_1 (x_1 - x_{1}^{\circ})^2 \\
    & &&V_2(x_1, x_2) &&= 0.5\cdot k_2 (x_2 - x_2^{\circ})^2  \\
    & &&V_3(x_1, x_2) &&= 0.5 \cdot \kappa \cdot 0.5 (C \tanh(\alpha(x_1 - x_{1}^{\circ})) +1.0) \cdot ((x_2 - x_1)-d)^2 \\
    & &&V &&= V_1 + V_2 + V_3  \\ \\
    &\textbf{System B} &&V_1(x_1) &&= 0.5 \cdot  k_1 (x_1 - x_1^{\circ})^2, 
\end{alignat*}
where $k_1 = 5.0, k_2 = 10.0, x_{1}^\circ = -2.0, x_{2}^\circ =2.0,, d=3.0, \kappa = 10.0, C = 0.5, \alpha = 2.0$.\\ \\

\textbf{Unbound State}
\begin{alignat*}{4}
    &\textbf{System A} \:\:\:\: &&V_1(x_1) &&=
    \begin{cases}
        0.5 \cdot  k (x_1 - x_{\text{l}})^2, \:\:\:\: \text{if}\ x_1 < x_{l} \\
        0.5 \cdot  k (x_1 - x_{\text{r}})^2, \:\:\:\: \text{if}\ x_1 > x_{\text{r}} \\
        0, \:\:\:\: \text{otherwise}
    \end{cases} \\
    & &&V_2(x_2) &&=
    \begin{cases}
        0.5 \cdot  k (x_2 - x_{l})^2, \:\:\:\: \text{if}\ x_2 < x_{l} \\
        0.5 \cdot  k (x_2 - x_{r})^2, \:\:\:\: \text{if}\ x_2 > x_{r} \\
        0, \:\:\:\: \text{otherwise}
    \end{cases} \\
    & &&V_3(x_1, x_2) &&= 0.5 \cdot  \kappa \cdot 0.5 (C \tanh(\alpha(x_1 - x_1^{\circ})) +1.0) \cdot ((x_2 - x_1)-d)^2\\
    & &&V &&= V_1 + V_2 + V_3\\ \\
    &\textbf{System B} \:\:\:\: &&V_1(x_1) &&=
    \begin{cases}
        0.5 \cdot  k (x_1 - x_{\text{l}})^2, \:\:\:\: \text{if}\ x_1 < x_{l} \\
        0.5 \cdot  k (x_1 - x_{\text{r}})^2, \:\:\:\: \text{if}\ x_1 > x_{\text{r}} \\
        0, \:\:\:\: \text{otherwise},
    \end{cases} 
\end{alignat*}
where $k = 10.0, x_{1}^\circ = -2.0, x_{l}  =3.0, x_{r} = 3.0, d=3.0, \kappa = 10.0, C = 0.5, \alpha = 2.0$.

\newpage
\FloatBarrier
\subsection{Training Results For Dual Flow Method.}
\label{apx_2d_hist}
\begin{figure}[ht]
\begin{center}
\includegraphics[width=.8\textwidth]{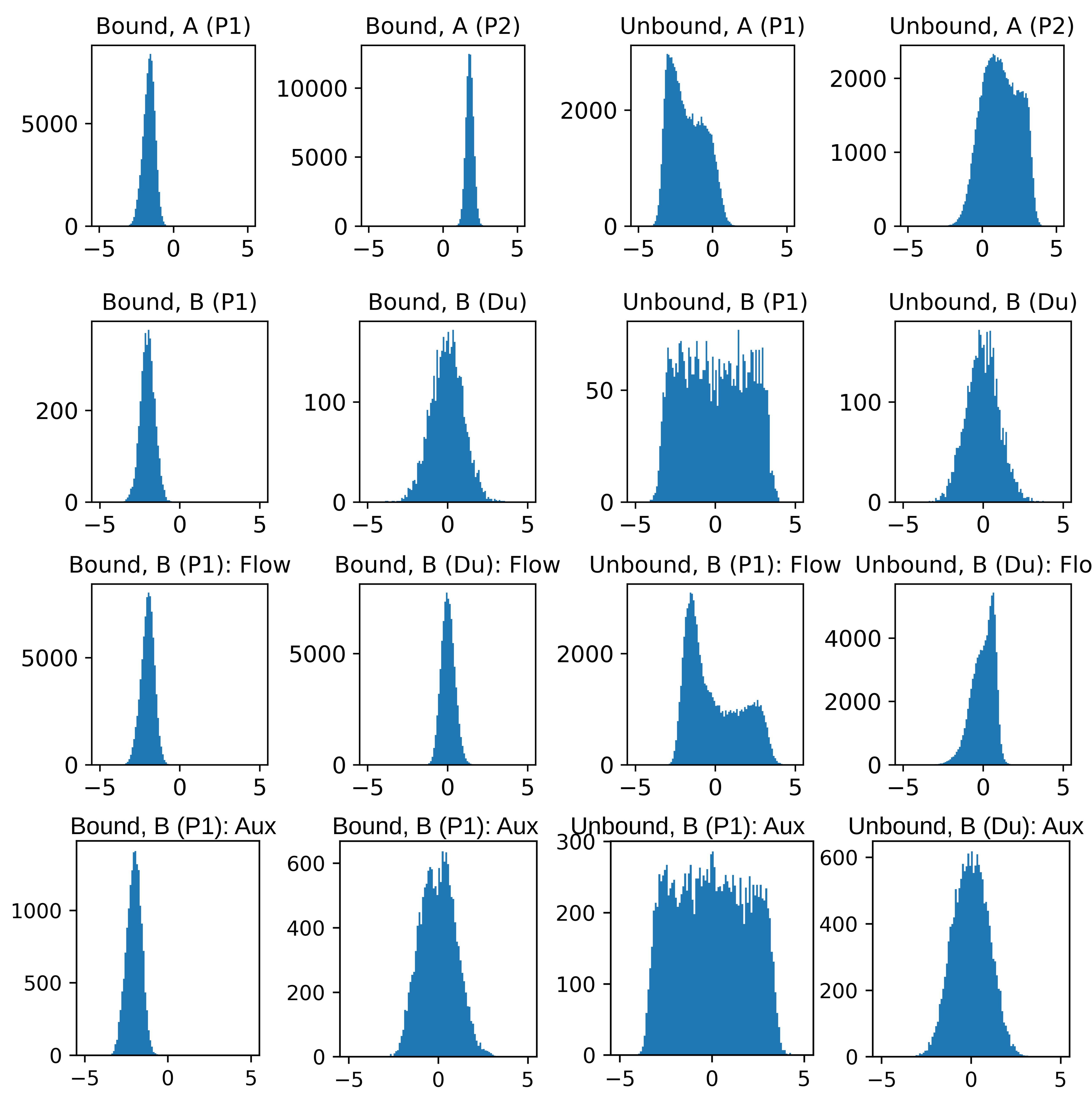}
\end{center}
\caption{Sampling histograms for the 2D coupled-particle system. System A is the two-particle system, System B is comprised of the first particle and an uncorrelated dummy atom that is defined by a Gaussian energy distribution with a mean of 0. The left two columns are for the bound state, the right two columns for the unbound state. The first row is the target System A, the second row target System B. The third row is the flow-generated distribution before using an auxiliary flow as prior. The final row shows correct training upon implementing the auxiliary flow. Improvements are especially more pronounced in histograms for the unbound state.}
\label{si_fig:2d_results}
\end{figure}

\subsection{Additional Information on the RFEP Method}
\label{apx_rfep}
Unique new atoms and their corresponding dummies receive coordinate placement by a reversible jump Monte Carlo geometry proposal engine to ensure generation of favorable molecular geometries that avoid steric clash or other errors in energy evaluation \citep{rufa_dominic_a_2022_6328265}. For the final stage, investigators have the option of using either equilibrium methods (e.g. replica exchange) or non-equilibrium methods (e.g. non-equilibrium switching) to estimate the free energy difference.

\subsection{Rational Quadratic Neural Spline Flows}
\label{apx_rqnsf}
Rational Quadratic Neural Spline Flows (RQNSFs) are a class of normalizing flow models that transform using rational quadratic spline functions (Equation \ref{si_eq_rqf}), or continuous functions mapped to a bound region (i.e. $[-B, B]$) for each DOF. The monotonic increase of rational quadratic functions ensure bijectivity in transform. The continuous first derivatives are piece-wise defined on intervals splitting this region. 
\begin{equation}
    \label{si_eq_rqf}
    z = f(x)= z_{0} + \frac{\left(z_{1}-z_{0}\right)\left[s \xi^2+d_{0}\xi\left(1-\xi\right)\right]}{s+\left[d_{1}+d_{0}-2s\right]\xi\left(1-\xi\right)},
\end{equation}
We note in the above equation that $x_{0(1)}, z_{0(1)},$ and $d_{0(1)}$ are the locations and derivatives at the left (right) interval boundaries, and $\xi = (x-x_{0})/(x_{1}-x_{0})$, $s = (z_{1}-z_{0})/(x_{1}-x_{0})$.

\begin{figure}[ht]
 \centering
 \includegraphics[width=.5\textwidth]{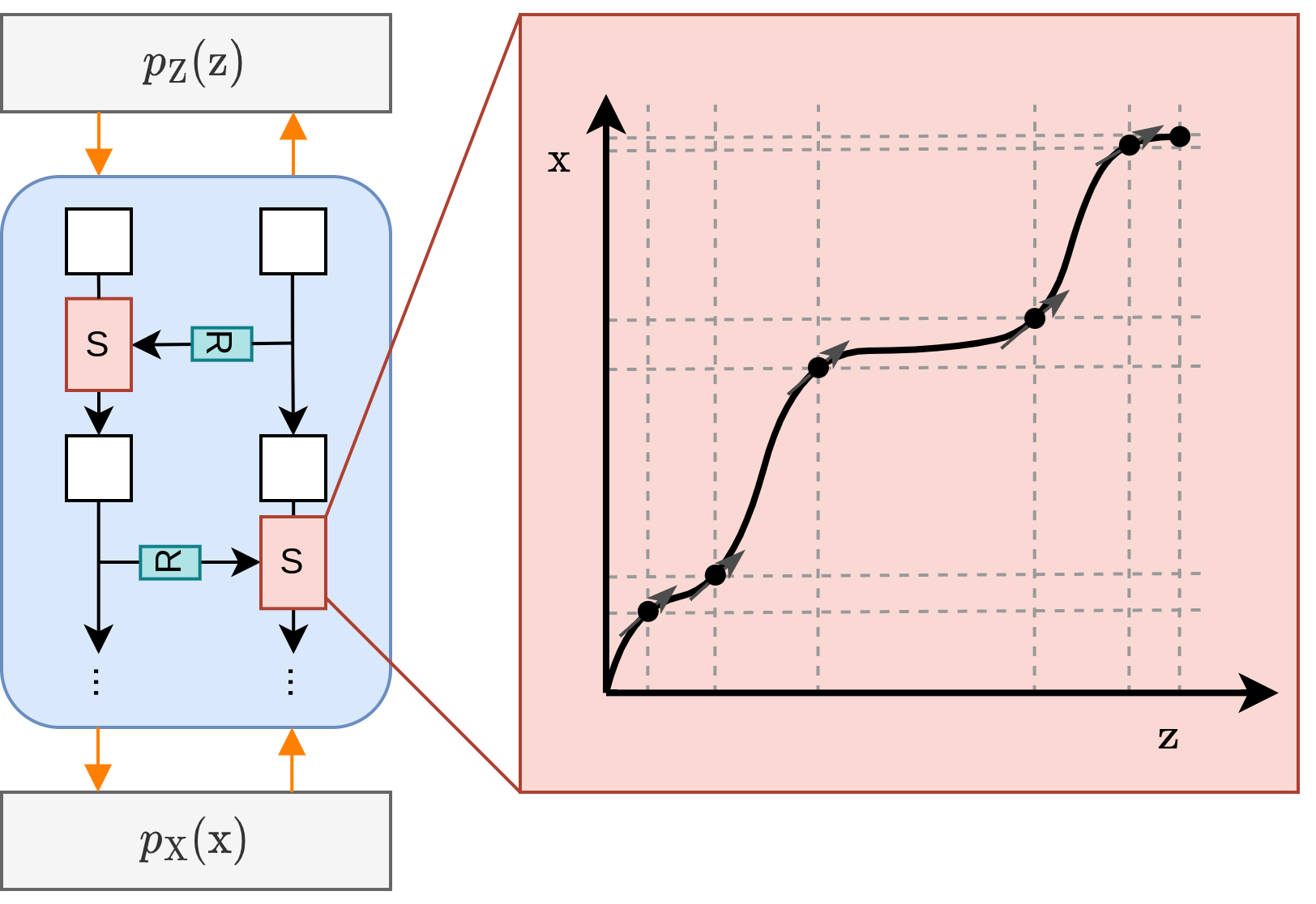}
 \caption{Scheme of a Rational Quadratic Spline function transformation. In our RQNSF implementation, a residual network is used to condition the spline coupling transformation.}
 \label{si_fig_rqnsf}
\end{figure}

We implemented RQNSF \cite{durkan2019neural} coupling layers in the mapping transformations of our architecture (Figure \ref{si_fig_rqnsf}) for a flow module expressive enough to capture the complex probability densities of systems that have high DOFs. Implementations are based on methods described in \citet{mahmoud2022accurate}.  

\subsection{Free Energy Estimation Methods}
\label{apx_fe_calculation}
In this section, we describe a few approaches to solvation free energy calculation for the sake of comparison with the TFEP approach. We define end states $A$ and $B$ with the respective Hamiltonians $\mathcal{H}_{\mathrm{A}}(\mathbf{q}, \mathbf{p} ; \lambda)$ and $\mathcal{H}_{\mathrm{B}}(\mathbf{q}, \mathbf{p} ; \lambda)$, where $\mathbf{q}$ and $\mathbf{p}$ represent positions and momenta of the system at given points in phase space and $\lambda$ the nonphysical coupling parameter. We then obtain the $\lambda$-dependent Hamiltonian as:
$$
\mathcal{H}(\mathbf{q}, \mathbf{p} ; \lambda)=f(\lambda) \mathcal{H}_{\mathrm{A}}(\mathbf{q}, \mathbf{p} ; \lambda)+g(\lambda) \mathcal{H}_{\mathrm{B}}(\mathbf{q}, \mathbf{p} ; \lambda).
$$

By convention, $\mathcal{H}=\mathcal{H}_{\mathrm{A}}$ at $\lambda=0$ and $\mathcal{H}=\mathcal{H}_{\mathrm{B}}$ at $\lambda=1$. The Hamiltonians are mixed using the functions $f(\lambda)$ and $g(\lambda)$. In the thermodynamic integration (TI) method, simulations are performed for a discrete set of $\lambda$ values and subsequently the free energy difference between $A$ and $B$ can be estimated by solving for $\Delta F=\int_{\lambda=0}^{\lambda=1}\left\langle\frac{\partial \mathcal{H}}{\partial \lambda}\right\rangle_\lambda \mathrm{d} \lambda$ by numerical quadrature approach. In the exponential averaging (EXP) or free energy perturbation (FEP) method \citep{zwanzig1954high}, $\lambda$-dependent Hamiltonians are similarly used to to estimate $\Delta F=-\frac{1}{\beta} \ln \left\langle\mathrm{e}^{-\beta\left[\mathcal{H}_{\mathrm{B}}(\mathbf{q} \mathbf{p} ; \lambda)-\mathcal{H}_A(\mathbf{q}, \mathbf{p} ; \lambda)\right]}\right\rangle_{\mathrm{A}}$

As a third alternative method, Bennett Acceptance Ratio (BAR) relies on bidirectional transfers to iteratively solve for the following: 
\begin{equation*}
    \left\langle\frac{1}{1+\frac{N_{\mathrm{A}}}{N_{\mathrm{B}}} \mathrm{e}^{\beta \Delta \mathcal{H}_{\mathrm{BA}}(\mathbf{q}, \mathbf{p})-\beta \Delta F}}\right\rangle_{\mathrm{A}} =\left\langle\frac{1}{1+\frac{N_{\mathrm{B}}}{N_{\mathrm{A}}} \mathrm{e}^{\beta \Delta \mathcal{H}_{\mathrm{AB}}(\mathbf{q}, \mathbf{p})+\beta \Delta F}}\right\rangle_{\mathrm{B}}.
\end{equation*}
Here, the respective numbers of statistically independent samples from $A$ and $B$ are denoted as $N_{\mathrm{A}}$ and $N_{\mathrm{B}}$ and the Hamiltonian differences as $\Delta \mathcal{H}_{\mathrm{BA}}(\mathbf{q}, \mathbf{p})=\mathcal{H}_{\mathrm{B}}(\mathbf{q}, \mathbf{p})-\mathcal{H}_{\mathrm{A}}(\mathbf{q}, \mathbf{p})=-\Delta \mathcal{H}_{\mathrm{AB}}(\mathbf{q}, \mathbf{p})$.
\newpage
\subsection{Ethane to Methanol Solvation Learning Curve}
\label{apx_solvation}

\begin{figure}[ht]
\begin{center}
\includegraphics[width=.8\textwidth]{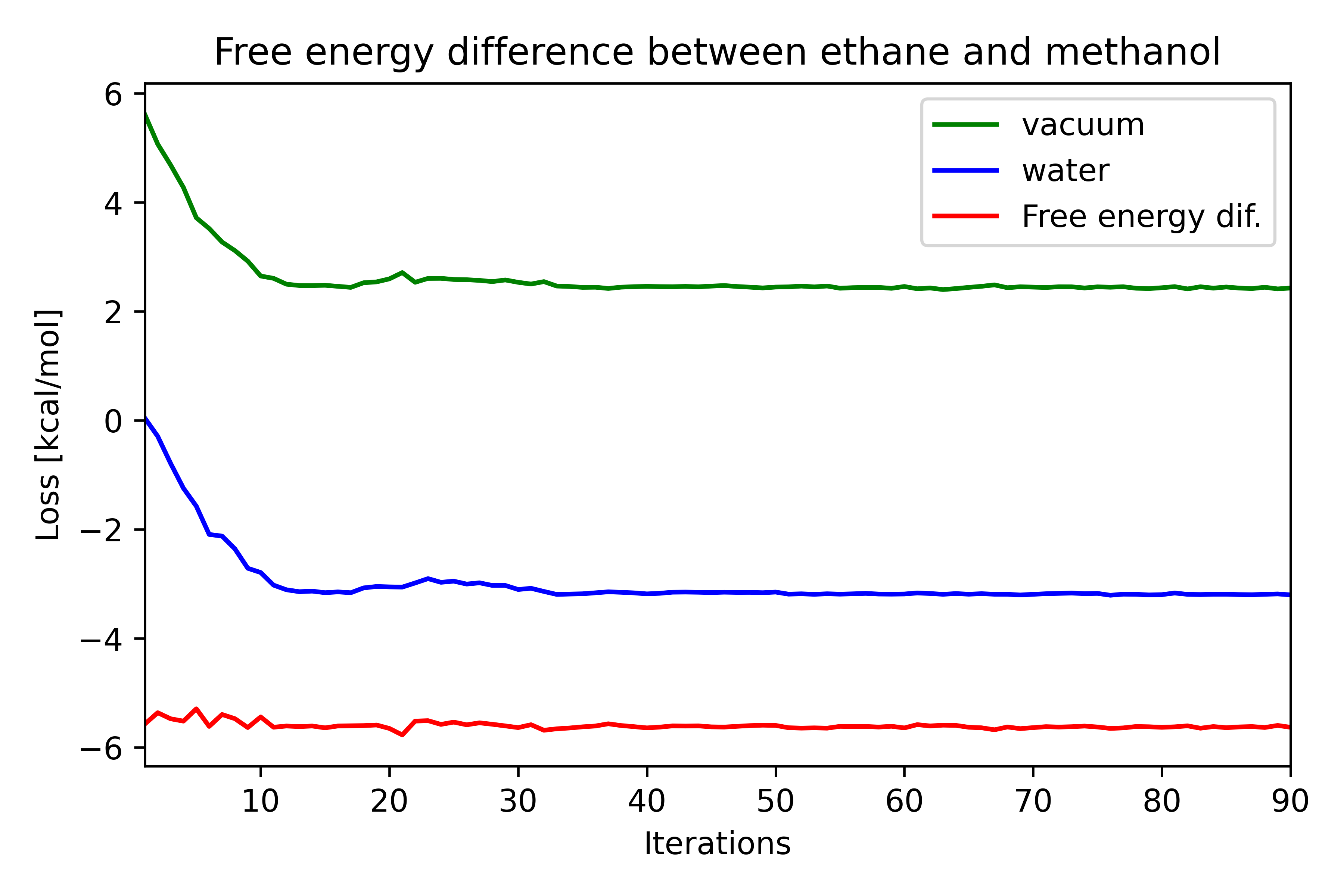}
\end{center}
\caption{Learning curve and free energy difference between ethane and methanol in water and vacuum.}
\label{si_fig:solvation_learning}
\end{figure}

\end{document}